\documentclass[%
 reprint,
superscriptaddress,
groupedaddress,
amsmath,amssymb,
aps,
prl,
floatfix
]{revtex4-1}

\usepackage{multirow}
\usepackage{graphicx}
\usepackage{dcolumn}
\usepackage{bm}
\usepackage[version=3]{mhchem}
\usepackage{nicefrac}

\usepackage{color}
\usepackage{dcolumn}
\usepackage{amssymb}
\usepackage{url}
\usepackage{hyperref}
\usepackage{xfrac}
\usepackage{tabularx}
\usepackage{xspace}
\usepackage{amsmath}
\usepackage[normalem]{ulem}
\usepackage{mathtools}

\def\be{\begin{equation}}
\def\ee{\end{equation}}
\def\bea{\begin{eqnarray}}
\def\eea{\end{eqnarray}}
\def\ba{\begin{array}}
\def\ea{\end{array}}

\def\MODELNAME{MLM\xspace}

\begin{document}
\title{ANOTHER EXACT GROUND STATE OF A 2D QUANTUM ANTIFERROMAGNET}
\author{Pratyay Ghosh}
\affiliation{Institut f\"ur Theoretische Physik und Astrophysik and W\"urzburg-Dresden Cluster of Excellence ct.qmat, Universit\"at W\"urzburg,
Am Hubland Campus S\"ud, W\"urzburg 97074, Germany}
\author{Tobias M\"uller}
\affiliation{Institut f\"ur Theoretische Physik und Astrophysik and W\"urzburg-Dresden Cluster of Excellence ct.qmat, Universit\"at W\"urzburg,
Am Hubland Campus S\"ud, W\"urzburg 97074, Germany}
\author{Ronny Thomale}
\affiliation{Institut f\"ur Theoretische Physik und Astrophysik and W\"urzburg-Dresden Cluster of Excellence ct.qmat, Universit\"at W\"urzburg,
Am Hubland Campus S\"ud, W\"urzburg 97074, Germany}

\begin{abstract}
We present the exact dimer ground state of a quantum antiferromagnet on the maple-leaf lattice. A coupling anisotropy for one of the three inequivalent nearest-neighbor bonds is sufficient to stabilize the dimer state. Together with the Shastry-Sutherland Hamiltonian, we show that this is the only other model with an exact dimer ground state for all two-dimensional lattices with uniform tilings. 
\end{abstract}

\maketitle

\textit{Introduction.} For decades, quantum antiferromagnets have been the center of condensed matter research~\cite{RevModPhys.25.58,RevModPhys.63.1,RevModPhys.78.17}. Frustrated magnetic couplings, combined with the non-commutativity of quantum spin operators, make it a challenging task to identify models which allow for an analytical understanding of their ground states. In particular, this applies to spatial dimensions greater than one, where integrability is more elusive, the Bethe ansatz, in general, does not apply, and conformal symmetry does not generate an extensive amount of conserved operators.

A first pivotal step in this direction was reached by Shastry and Sutherland in 1981~\cite{Shastry1981} where, in continuation of a spin chain model by Majumdar and Ghosh~\cite{Majumdar1969}, the first quantum antiferromagnet with uniform tilings was found to exhibit an exact dimer ground state. While a significant fraction of the research activity subsequently shifted to topologically ordered exact ground states of quantum antiferromagnets such as chiral spin liquids~\cite{PhysRevLett.59.2095,PhysRevLett.99.097202}, valence bond liquids~\cite{PhysRevB.35.8865,PhysRevLett.86.1881}, or the $Z_2$ spin liquid realized in the Kitaev model~\cite{PhysRevB.44.2664,PhysRevB.65.165113,KITAEV20062}, the Shastry-Sutherland model (SSM) has prevailed as an important crystallization point for discoveries and theoretical developments in quantum antiferromagnets.     

\begin{figure}
\includegraphics[width=0.95\columnwidth]{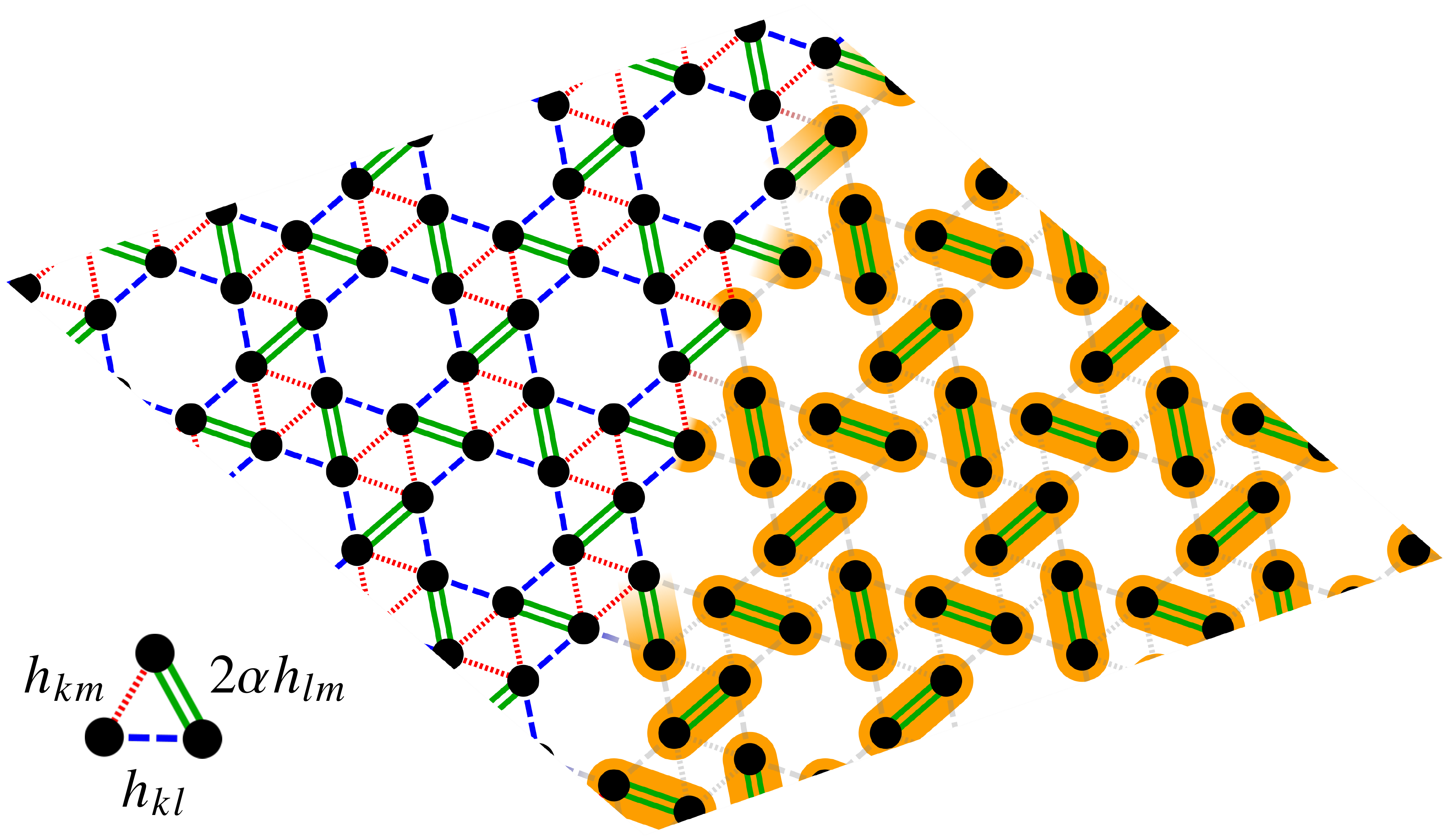}
\caption{Maple-leaf lattice (left) and the dimer eigenstate of the Hamiltonian in \eqref{eq-ham} (right). According to \eqref{eq-ham} there are three inequivalent nearest neighbor couplings denoted red (dotted), blue (dashed), and green (double line). The singlets (yellow ellipses) reside on the green bonds. As the red and blue couplings do not contribute to the energy, it is an eigenstate of \eqref{eq-ham}.} \label{fig-lat}
\end{figure}

In this Letter, we propose a spin model defined on the maple-leaf lattice. The maple-leaf tiling is called snub trihexagonal tiling ($p6$ space group), where four triangles and one hexagon surround each site of the lattice (Fig.~\ref{fig-lat}). This corresponds to a $1/7$-site ($1/6$-bond) depleted version of the triangular lattice with coordination number $5$~\cite{Betts1995}. We find that the phase diagram of our model hosts an exact dimer ground state, and show that our Hamiltonian, aside from the SSM, is the only other such 2D model with uniform tilings.   

\textit{Model.} Our Hamiltonian is given by 
\be\label{eq-ham}
\mathcal{H}=\sum_{\mathclap{\langle kl\rangle}} h_{kl}+\sum_{\mathclap{\langle km\rangle}}h_{km}+2\alpha\sum_{\mathclap{\langle lm\rangle}}h_{lm}+B\sum_{i}S_{i}^z.
\ee
Aside from a Zeeman term, there are separate summations over the three inequivalent nearest neighbor bonds $kl$, $km$, and $lm$ on the maple leaf lattice denoted blue (dashed), red (dotted), and green (double line), respectively (Fig.~\ref{fig-lat}). $h_{ij}$ represents the XXZ-type spin exchange interaction ($J_z,J_\perp>0$) 
\be
h_{ij}=J_zS_{i}^zS_{j}^z+J_{\perp}(S_{i}^xS_{j}^x+S_{i}^yS_{j}^y)
\ee
between the sites $i$ and $j$, where $S^{\mu}_i$ denotes the $\mu=x,y,z$ component of the $\mathfrak{su}(2)$ spin operator acting on the spin-$S$ representation on site $i$. We label~\eqref{eq-ham} as the maple-leaf model (MLM). The Heisenberg model on the maple-leaf lattice has been studied previously~\cite{Misguich1999,Schmalfuss2002,Farnell2011,Makuta2021}, where the exact dimer state is argued to be an eigenstate~\cite{Misguich1999} and some numerical indication of a pronounced dimerization propensity has been found~\cite{Farnell2011}.

\textit{Ground State Analysis.} To achieve an exact solution of the \MODELNAME, we first rewrite the bond summations in~\eqref{eq-ham} as the sum over interacting spins on triangles
\be\label{eq-hamil-tri}
\mathcal{H}=\sum\left(\begin{gathered}\includegraphics[scale=0.14]{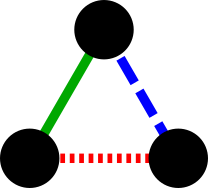}\end{gathered}+\begin{gathered}\includegraphics[scale=0.14]{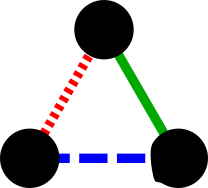}\end{gathered}+\begin{gathered}\includegraphics[scale=0.14]{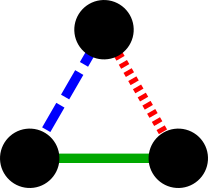}\end{gathered}+\begin{gathered}\includegraphics[scale=0.14]{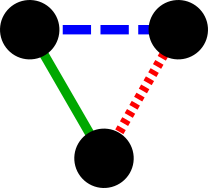}\end{gathered}+\begin{gathered}\includegraphics[scale=0.14]{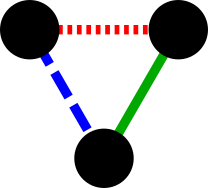}\end{gathered}+\begin{gathered}\includegraphics[scale=0.14]{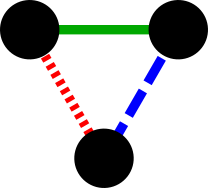}\end{gathered}\right).
\ee
For this, we have split the interaction $2\alpha h_{lm}$ on the double green bonds in Fig.~\ref{fig-lat} to construct two adjacent triangles sharing this bond. These triangles are formed by pairwise differently colored bond interactions  
\be\label{eq-ham2}
h_{\triangle}=h_{kl}+h_{km}+\alpha h_{lm}+\frac{B}{2}(S_l^z+S_m^z).
\ee
From the triangular decomposition, \eqref{eq-ham} turns out to be a frustration-free model~\cite{Majumdar1969,Shastry1981,AKLT,Klein_1982}, in the sense that the ground state minimizes the energy of each $h_\triangle$ individually. Note that if there are $N$ sites for the maple-leaf lattice, there are $N$ such tri-colored triangles and therefore $N/2$ green bonds. 

For large $\alpha$, the ground state of a single tri-colored triangle is a singlet of the spins forming the green bond, which is strongest among the three types of nearest neighbor bonds, denoted by $\big| [lm] \big\rangle$. Moreover, $(h_{kl}+h_{km})\big| [lm] \big\rangle=0$, leading to the first two terms in~\eqref{eq-ham2} not to contribute to the energy of the triangle. We construct the tensor product state 
\be
|\psi\rangle=\underset{\mathclap{\langle lm \rangle}}{\bigotimes} \big| [lm] \big\rangle, \label{eigen}
\ee
which covers the entire spin lattice (see Fig.\ref{fig-lat}). The first two terms of~\eqref{eq-ham} do not renormalize $|\psi\rangle$, thus making \eqref{eigen} an exact eigenstate of the \MODELNAME. The corresponding energy density, which is solely determined by the interactions on the green bonds, is given by 
\be
E/N= -\alpha \frac{S(S+1)}{3}(J_z+2J_\perp).\label{energ}
\ee
From the variational principle, if $e_\triangle$ is the ground state energy of the individual triangles~\eqref{eq-ham2}, then~\eqref{energ} serves as an upper bound for the ground state energy density of the system, i.e., $E_0/N\geq e_\triangle$. The equality $E_0/N=e_\triangle$ should hold when $\alpha$ is greater than a lower bound $\alpha_{b1}$, where the dimer state, $|\psi\rangle$, becomes the exact ground state of the MLM. For spin-$1/2$, this bound is given by $\alpha_{b1}=\frac{(B+J_z)+\sqrt{(B+J_z)^2+4J_\perp(J_z+J_\perp)}}{2(J_z+J_\perp)}$, and can likewise be obtained for other spin-$S$ representations~\footnote{See Supplemental Material at [URL will be inserted by publisher].}. Note that while \eqref{eigen} ceases to be the ground state of the MLM for $\alpha < \alpha_{b1}$ from a variational principle, it might still be the exact ground state. In any case, \eqref{eigen} still is an exact eigenstate of the MLM, bearing some similarity to the motif of scar states in the context of many-body localization~\cite{scarMaple}.

In order to allow for an exact dimer ground state in accordance with the triangle decomposition explained above, any model needs to satisfy two conditions: (i) every bond must be part of a triangle where at least one bond is symmetry-inequivalent and (ii) this bond must be shared between two triangles. For \MODELNAME, the green bonds are not related by symmetry to the other bonds, which satisfies (i), and are shared between two triangles, which satisfies (ii). Note that, to meet (i), we intentionally avoid a triangle decomposition involving red bond only triangles, but arrange them in three tri-colored triangles accordingly. Further analyzing the geometric restrictions of uniformly tiling Euclidean space~\cite{Grunbaum1977,Grunbaum1986}, the exact dimer state construction above confines us to the lattices with coordination number $5$ made with the tiles of vertex configuration $3.3.3.3.6$ or $3.3.4.3.4$, where the sequence of numbers represents the number of sides of the faces around the vertex~\cite{Note1}. $1$-uniform tilings of the former generate the maple-leaf lattice, while the latter yields the Shastry-Sutherland model (SSM), where the existence of an exact dimer state was first found~\cite{Shastry1981}. 
These two tiles can also produce other lattices with $k$-uniform tilings for $k\ge 2$, but either condition (i) or (ii) is violated in all those cases. Thus, the \MODELNAME and SSM exhaust the list of all uniform 2D lattices that can host an exact dimer state \eqref{eigen}. Except for \MODELNAME and SSM, any alternative 2D model with an exact dimer state is either defined on a lattice with non-uniform tilings or includes further neighbor couplings~\cite{Siddharthan1999,SchmidtExact}. For the remainder part of this Letter where we focus on the analysis of the coupling anisotropy $\alpha$ in the MLM, we confine ourselves to \eqref{eq-ham} with $J_z=J_\perp=1$ and $B=0$.
 
\textit{Classical Limit.} 
Setting $S \to \infty$~\cite{Luttinger1946,Kaplan2007}, the \MODELNAME for $\alpha\leq1$ yields local $120^{\circ}$-order on individual red triangles, with a non-local spin canting induced by the blue and the green bonds~\cite{Schulenburg2000, Schmalfuss2002,Farnell2011}, which we denote as \emph{canted} $120^{\circ}$ (c$120^{\circ}$) order in the following. The energy is given by \be
E_c/N = -\frac{1}{2} + \cos(\Phi) + \alpha \sin\left(\Phi - \frac{\pi}{6}\right),\label{eq:Eclassicalphi}
\ee
where $\Phi = \pi - \cos^{-1}\left(\frac{2-\alpha}{2\sqrt{\alpha^2-\alpha+1}}\right)$ parametrizes the canting between spins across the blue bonds (Fig. 2).     

The canting can be reconciled from the limit $\alpha = 0$, where the frustrated red triangles will assume a $120^\circ$ order, while blue bonds only provide antiparallel orientation of neighboring spins without introducing any additional frustration. For $\alpha > 0$, the green bonds start to contest the antiferromagnetic ordering on the blue bonds, introducing the canting between neighboring spins on different red triangles. In the particular case of $\alpha = 1$, the \MODELNAME produces a uniform $120^{\circ}$-order akin to the Heisenberg antiferromagnet on a triangular lattice. Similar to the exact dimer ground state analysis above, one can understand this as one splits the green bonds, leading to isolated triangular motifs of three tri-colored triangles, where the equality of all bond couplings establishes the uniform $120^{\circ}$-order.  
\begin{figure}[b]
    \centering
    \includegraphics[width=0.7\columnwidth]{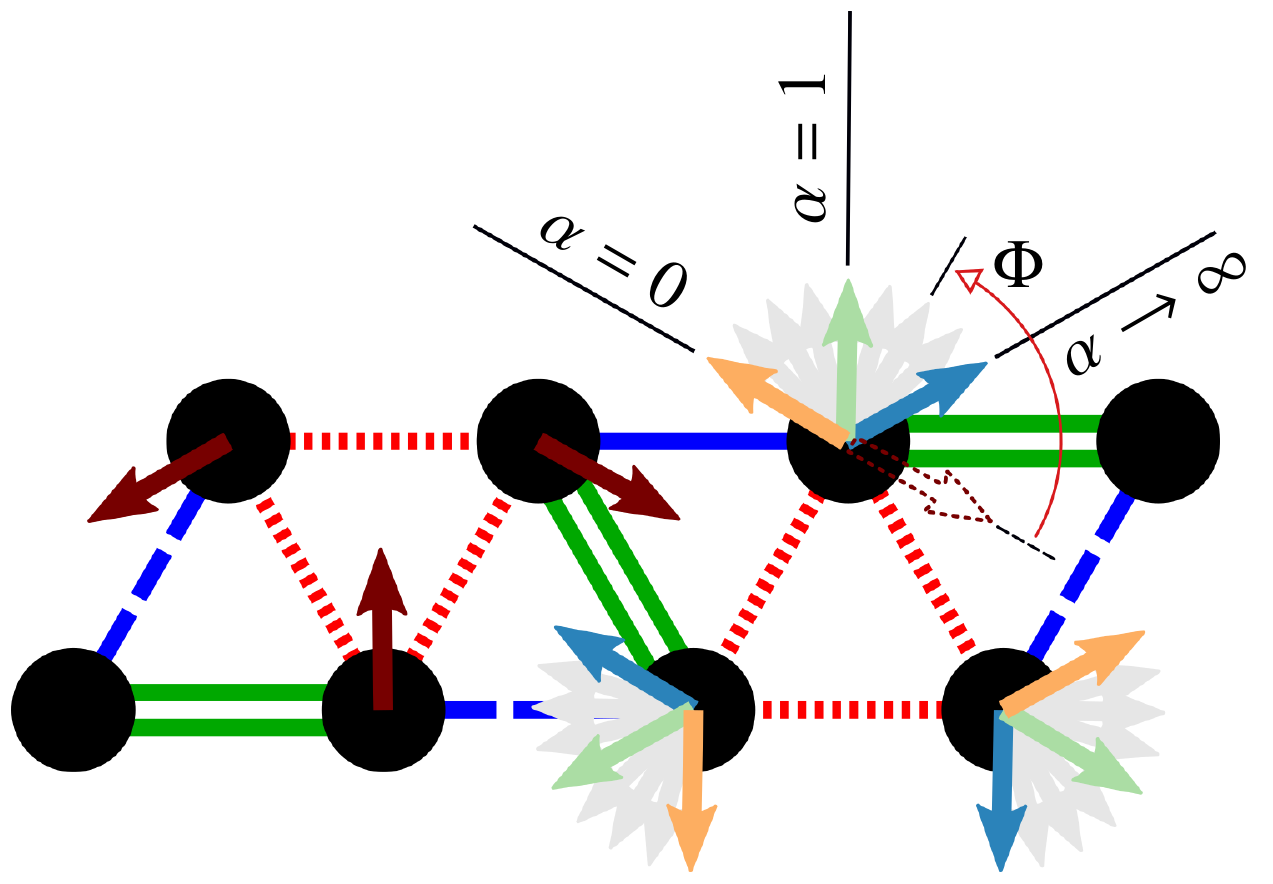}
    \caption{Canted $120^{\circ}$-order: spins on red triangles (brown) show $120^{\circ}$-order, while those connected by blue bonds are canted by an angle $\Phi(\alpha)$. As a function of $\alpha$, spins across blue bonds are antiparallel for $\alpha=0$ (orange), uniform $120^{\circ}$-order appears for $\alpha=1$ (light green), and decoupled green dimer bonds are formed for $\alpha\to\infty$ (light blue).}
    \label{fig:my_label}
\end{figure}
For $\alpha>1$, aside from the trivial decoupled dimer state limit for $\alpha \to \infty$, a large-$N$ analysis~\cite{Isakov2004} ceases to give a unique ground state. There, as one effectively removes the spin normalization constraint, we find a subextensive degeneracy of ground states ~\cite{Note1}. In contrast to the c$120^{\circ}$-state for $\alpha<1$, this manifold cannot form a normalized spin state. In accordance with earlier studies~\cite{Farnell2011}, for normalized spins, we do not find any lower energy state than the c$120^{\circ}$ state. 

\begin{figure}
\includegraphics[width=0.95\columnwidth]{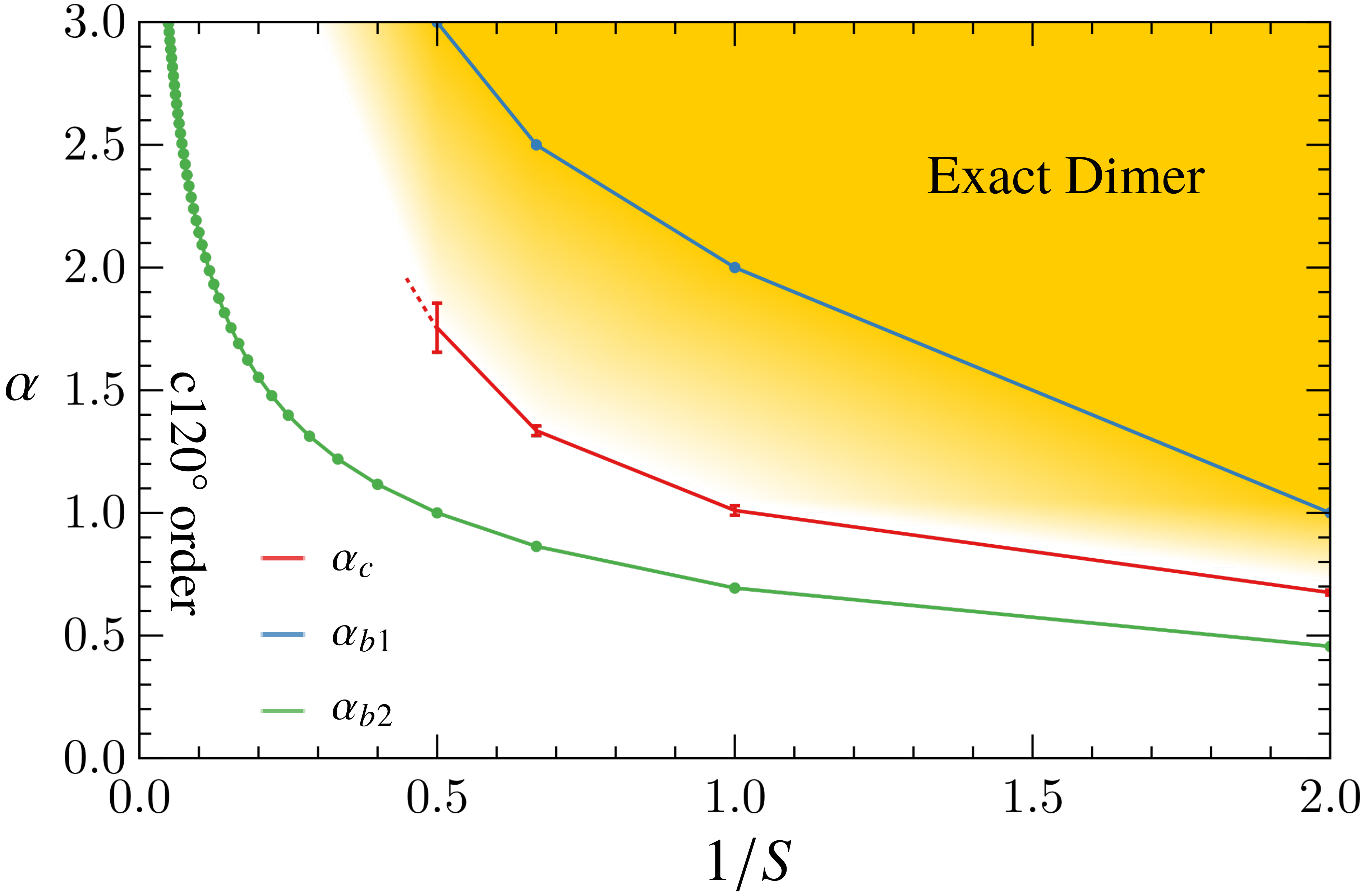}
\caption{Dimer phase diagram of \MODELNAME as function of bond coupling $\alpha$ and spin $S$. $\alpha_{b1}$ denotes a lower bound above which the \MODELNAME is exactly solvable. $\alpha_{b2}$ is an upper bound of $\alpha$ below which c$120^\circ$ serves as a better lower bound of energy than \eqref{energ}. $\alpha_c$ is the numerically obtained critical value of $\alpha$ above which \eqref{eigen} stabilizes.} \label{fig-QPD}
\end{figure}

The particular source of frustration in the \MODELNAME plays a pivotal role in preventing the system from achieving a global energy minimum by simultaneously minimizing the individual local energies of the tri-colored triangles, and thereby forbids, apart from the special point $\alpha=1$, the stabilization of a commensurate magnetic ordering. Instead, our observed canted $120^{\circ}$ order bears similarities to frustration-driven \emph{block-spiral magnetism} found in one-dimensional strongly interacting itinerant fermion models away from half-filling (see Ref.~\cite{Herbrych2020} and references therein), with $120^{\circ}$-ordered triangles spiraling with an $\alpha$-dependent pitch. To our knowledge, the \MODELNAME is the only example where such a state is realized in either two spatial dimensions or within a local spin model.

\textit{Dimer Phase Diagram.} Together with $\alpha_{b1}$ from the exact dimer ground state analysis, we complement our analysis of the MLM through a density matrix renormalization group (DMRG) study in order to obtain an estimate of the critical $\alpha_c$, such that for $\alpha\geq\alpha_c$, \eqref{eigen} is the exact ground state. The DMRG calculations are performed on a $108$ site cluster~\cite{Note1} using the ITensor library~\cite{Fishman2020}. Combined, the dimer phase diagram of the MLM is presented in Fig.~\ref{fig-QPD}, where we only show those results for different values of $S$ where the DMRG results are converged. As expected, $\alpha_c$ turns out to be smaller than $\alpha_{b1}$ which takes the values $1$ for $S=1/2$ and $1+S$ for $S\ge 1$. Even though this is beyond the numerical range of investigation, we expect $\alpha_c$ to converge asymptotically against $\alpha_{b1}$ for increasing $S$. For the sake of an additional consistency check, we evaluate  $\alpha_{b2}$ below which the c$120^{\circ}$ state serves as a better lower bound for the energy than the dimer state~\eqref{eigen}, by comparing $e_\triangle$ and the c$120^{\circ}$ energy given in~\eqref{eq:Eclassicalphi} rescaled by spin $S$. We find $\alpha_{b2}=\frac{S}{2S+1}\left(\frac{1}{2}+\sqrt{\frac{3S+2}{2}}\right)<\alpha_c$. Together, the yellow regime in Fig.~\ref{fig-QPD} highlights the parameter space in which we find a dimer ground state. Note that, for our dimer phase diagram, we have only focused on the stability of the nearest neighbor dimer state \eqref{eigen}. Furthermore, in order to specify an ordered state out of which the dimerized regime might evolve, the classical limit has provided us with a candidate c$120^{\circ}$ order. Our analysis does not exclude the possibility that the \MODELNAME can host other exotic quantum disordered phases due to its high frustration for $\alpha<\alpha_c$.  It highlights the nature of the phase transition into the dimerized phase, and possibly additional phase transitions featuring deconfined criticality~\cite{deconfined}, as a future problem.

It is insightful to compare our dimer phase diagram Fig.~\ref{fig-QPD} with the related scenario in the SSM, for which we have already chosen the appropriate parametrization of \eqref{eq-ham} in hindsight. We find that the \MODELNAME is even more frustrated than the SSM, leading to a stable dimer ground state extending to lower $\alpha$: For the isotropic case, the SSM is known to yield $\alpha_c^{\text{SSM}}\approx0.74$ \cite{Corboz2013,Lee2019}. Small cluster results from exact diagonalization and the Coupled Cluster Method on a related model to MLM suggest $\alpha_c$ to be $0.725$~\cite{Farnell2011}. Our larger cluster DMRG calculations, however, indicate an even substantially smaller value $\alpha_c\approx0.675$. The \MODELNAME thus sets a new bar of frustration for a 2D model with uniform tilings where an exact dimer ground state appears. In  comparison to the SSM, this roots in the additional frustration emanating from the red triangles (Fig.~1). 

\textit{Conclusions and Outlook.} 
We have introduced the exact dimer ground state solution of the \MODELNAME. Aside from the SSM, we find that it is the only other such model in two spatial dimensions with uniform tilings exhibiting such a property. In comparison to the SSM, we find a significantly larger region of stability of the dimer ground state, which is a consequence of the enhanced frustration inherent to the \MODELNAME. Magnetic ordering phenomena in the MLM likewise promise highly exotic behaviour. This already becomes apparent from the c$120^{\circ}$ order in the classical limit, where the canting continuously adjusts to the given bond anisotropy of interactions $\alpha$.

The \MODELNAME opens up several further theoretical and experimental explorations. First, the stability of the dimer state \eqref{eigen} encourages further investigation within an extended parameter space involving bond and spin exchange anisotropies in~\eqref{eq-ham}. Second, by allowing for finite $B$ in \eqref{eq-ham}, the \MODELNAME is expected to show a series of magnetization plateaus and triplet bound states, which promises a phenomenology as rich as the SSM~\cite{Momoi2000,Fukumoto2000,Miyahara2003,Dorier2008,Corboz2014}. The exact dimer state has a 3-dimer unit cell, whose minimal excitation is a single triplet. Therefore, from this elementary estimate, $1/3$ and $2/3$ magnetization plateaus will most certainly appear, complemented by further plateaus resulting from a triplet bound state hierarchy, which is typically located at fractions that are rational multiples of $1/3$. This reasoning would already explain the early findings from preliminary exact diagonalization studies in Ref.~\cite{Farnell2011}.

There are multiple material realizations of a spin system on the maple-leaf lattice~\cite{Cave2006,Fennel2011,Aliev2012,Haraguchi2018,Haraguchi2021}. For the \MODELNAME, \ce{MgMn3O7.3H2O} can be a promising candidate. In general, similar to how the SSM and its experimental realizations have evolved in the past decades, we consider it a fruitful enterprise to reexamine existing maple-leaf compounds and their derivative material families to find new compounds that could host the \MODELNAME dimer state. This might be further facilitated by the record high frustration of the \MODELNAME, as the required minimal bond anisotropy $2\alpha$ is smaller than for the SSM.

\textit{Acknowledgments.} We thank M.~Greiter, Y.~Iqbal, M.~Klett, B.~Kumar, R. Moessner, and J.~Reuther for illuminating discussions, and, in particular, Y.~Iqbal for drawing our attention to the maple-leaf lattice.
The work in W\"urzburg is supported by the Deutsche Forschungsgemeinschaft (DFG, German Research Foundation) through Project-ID 258499086-SFB 1170 and the Würzburg-Dresden Cluster of Excellence on Complexity and Topology in Quantum Matter – ct.qmat Project-ID 390858490-EXC 2147.

\bibliography{Refs}

\end{document}


\title{Supplementary Material: ANOTHER EXACT GROUND STATE OF A 2D QUANTUM ANTIFERROMAGNET}
\author{Pratyay Ghosh}
\author{Tobias M\"uller}
\author{Ronny Thomale}
\affiliation{Institut f\"ur Theoretische Physik und Astrophysik and W\"urzburg-Dresden Cluster of Excellence ct.qmat, Universit\"at W\"urzburg,
Am Hubland Campus S\"ud, W\"urzburg 97074, Germany}

\maketitle

\section{Lattice conventions}
\begin{figure}
    \centering
    \includegraphics[width=0.25\textwidth]{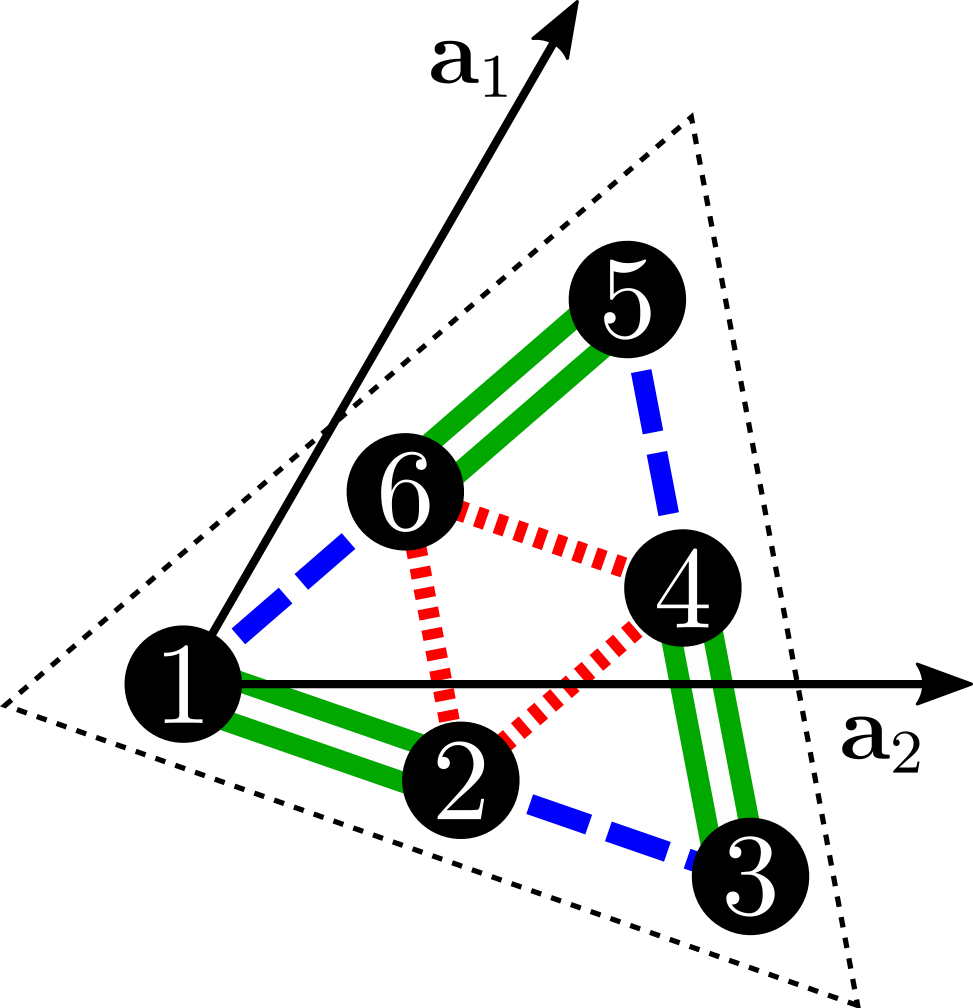}
    \caption{Unitcell of the maple-leaf lattice, showing the numbering convention of basis sites and the primitve vectors $\mathbf{a}_1$ and $\mathbf{a}_2$ adopted.}
    \label{fig:unitcell_supp}
\end{figure}

Here we give the conventions used for constructing the maple-leaf lattice. We set the nearest-neighbor distance to unity. The Bravais lattice vectors shown in Fig.~\ref{fig:unitcell_supp} are then given by
\begin{align}
    \mathbf{a}_1  &= \frac{\sqrt{7}}{2}\begin{pmatrix}1\\\sqrt{3}\end{pmatrix} & 
    \mathbf{a}_2 &= \sqrt{7}\begin{pmatrix}1\\0\end{pmatrix}.
\end{align}

The six basis sites are located at
\begin{equation}
\begin{aligned}
    \mathbf{b}_1  &= \begin{pmatrix}0\\0\end{pmatrix} & 
    \mathbf{b}_2 &= \frac{1}{2\sqrt{7}}\begin{pmatrix}5\\-\sqrt{3}\end{pmatrix} & 
    \mathbf{b}_3  &= \frac{1}{\sqrt{7}}\begin{pmatrix}5\\-\sqrt{3}\end{pmatrix}\\
    \mathbf{b}_4  &= \frac{1}{2\sqrt{7}}\begin{pmatrix}9\\\sqrt{3}\end{pmatrix} & 
    \mathbf{b}_5  &= \frac{1}{\sqrt{7}}\begin{pmatrix}4\\2\sqrt{3}\end{pmatrix} & 
    \mathbf{b}_6 &= \frac{1}{2\sqrt{7}}\begin{pmatrix}4\\2\sqrt{3}\end{pmatrix},
    \end{aligned}
\end{equation}
where the numbering of the sites is given in Fig.~\ref{fig:unitcell_supp}.

\section{Parametrization of the canted $120^{\circ}$ order}
For reference, we give a parametrization of the full c120$^\circ$ state. The spin directions on the respective basis points, up to a possible global spin rotation and reflection, given by
\begin{equation}
    \begin{aligned}
    \mathbf{S}_1  &= \begin{pmatrix}\cos(\mathbf{x}\cdot\mathbf{q})\\[0.5em] \sin(\mathbf{x}\cdot\mathbf{q})\\[0.5em] 0\end{pmatrix} & 
    \mathbf{S}_2 &= \begin{pmatrix}\cos(\mathbf{x}\cdot\mathbf{q}+\frac{2\pi}{3}+ \Phi(\alpha))\\[0.5em] \sin(\mathbf{x}\cdot\mathbf{q}+\frac{2\pi}{3}+ \Phi(\alpha))\\[0.5em] 0\end{pmatrix} & 
    \mathbf{S}_3  &= \begin{pmatrix}\cos(\mathbf{x}\cdot\mathbf{q}+\frac{2\pi}{3})\\[0.5em]\sin(\mathbf{x}\cdot\mathbf{q}+\frac{2\pi}{3})\\[0.5em]0\end{pmatrix}\\
    \mathbf{S}_4  &= \begin{pmatrix}\cos(\mathbf{x}\cdot\mathbf{q}-\frac{2\pi}{3}+ \Phi(\alpha))\\[0.5em]\sin(\mathbf{x}\cdot\mathbf{q}-\frac{2\pi}{3}+ \Phi(\alpha))\\[0.5em]0\end{pmatrix} & 
    \mathbf{S}_5  &= \begin{pmatrix}\cos(\mathbf{x}\cdot\mathbf{q}-\frac{2\pi}{3})\\[0.5em]\sin(\mathbf{x}\cdot\mathbf{q}-\frac{2\pi}{3})\\[0.5em]0\end{pmatrix} & 
    \mathbf{S}_6 &= \begin{pmatrix}\cos(\mathbf{x}\cdot\mathbf{q}+ \Phi(\alpha))\\[0.5em]\sin(\mathbf{x}\cdot\mathbf{q}+ \Phi(\alpha))\\[0.5em]0\end{pmatrix},
    \end{aligned}
\end{equation}

where $\mathbf{q} = (4\pi/(3 \sqrt{7}), 0)$ and $\mathbf{x}$ designates the location of the unitcell, i.e. it is the same for all basis points. The angle $\Phi$ is given in Eq.~(11) of the main text.

\section{$\alpha_{b1}$ for some limiting cases}
For $S\geq1$, \be\alpha_{b1}=1+S\ee for the isotropic case ($J_z=J_\perp=1$) without external field. This result has also been used in the main text.

In the presence of external magnetic fields $\alpha_{b1}$ is modified and for $S=1$ and $3/2$ as follows.
\bea
\alpha_{b1}(S=1)&=&2+\frac{B^2}{6}+\mathcal{O}(B^3)\\
\alpha_{b1}(S=3/2)&=&\frac{5}{2}+\frac{B}{6}+\frac{5B^2}{54}+\mathcal{O}(B^3).
\eea 
The result for $S=1/2$ can be easily derived from the expression of $\alpha_{b1}$ given in the \textit{Ground State Analysis} section of the main text. 
\section{$O(N\to \infty)$ results for $\alpha>1$}
\begin{figure}
    \centering
    \includegraphics[width=0.8\textwidth]{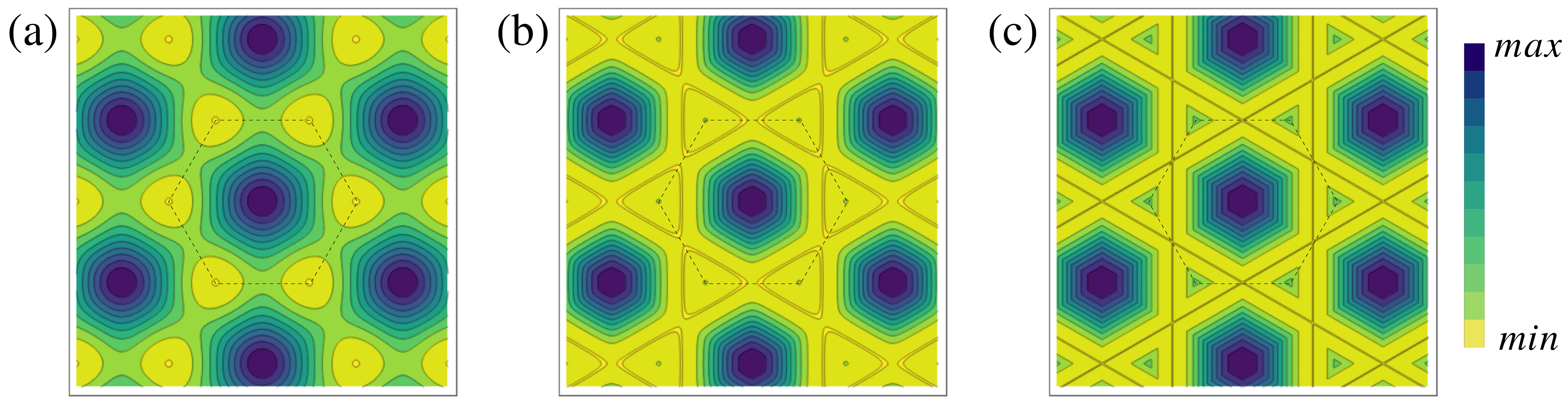}
    \caption{Luttinger-Tisza energy landscape of the maple-leaf lattice for (a) $\alpha=0$, (b) $\alpha=2.5$ and (c) $\alpha=100$. The black dashed hexagons shows the Brillouin zone.}
    \label{fig:LT_energies}
\end{figure}
In the parameter region $\alpha \leq 1$, the classical model with spin length $S \to \infty$, we use the Luttinger-Tisza method~\cite{Luttinger1946, Kaplan2007} to determine the classical ground state. In this formalism, the spin length constraint of the classical model is only enforced on average, which is equivalent to enlarging the classical spin rotation symmetry from $O(3)$ to $O(N)$ in the limit $N \to \infty$ \cite{Isakov2004}. For $\alpha \leq 1$, the Luttinger-Tisza states with minimal energy are located at the Brillouin zone corners (see Fig.~\ref{fig:LT_energies}(a)) and correspond to the c120$^\circ$ spin state. As in this Luttinger-Tisza state, the spins are normalized; it represents the exact classical ground state.

The situation changes for $\alpha>1$: The Luttinger-Tisza energy minima start developing into triangular-shaped valleys around the corners, which become more prominent with increasing $\alpha$ (see Fig.~\ref{fig:LT_energies}(b) and (c)). For $\alpha \to \infty$ they merge into a Kagome-like net of energy minima in reciprocal space.
These subextensively degenerate states no longer correspond to normalizable states, i.e., another ordering vector can govern the classical ground state. Indeed, we have not found any classical state in this parameter region, which has lower energy than the c120$^\circ$-order.

However, the degeneracy of the Luttinger-Tisza eigenstates is already an indicator that fluctuations, either thermal or quantum in nature, can quickly destroy the classical order, as seen in this model, where the exact dimer state takes over in the quantum limit.

\section{Exact dimer ground state in 2D: Analysis from lattice geometries}\label{supp:latt}
This part of the supplement discusses the exact ground states in nearest-neighbour 2D (and partly 3D) lattices with uniform semi-regular tilings.

\textbf{Definition 1:} 
Uniform Lattice -- A lattice constructed of uniform semi-regular tilings, i.e., tilings made of regular polygons. There are $21$ ways to fit regular polygons around a vertex on the Euclidean plane. However, the number lattices made of these vertex configurations are infinitely many~\cite{Grunbaum1977,Grunbaum1986}. 
\begin{figure}
    \centering
    \includegraphics[width=0.95\textwidth]{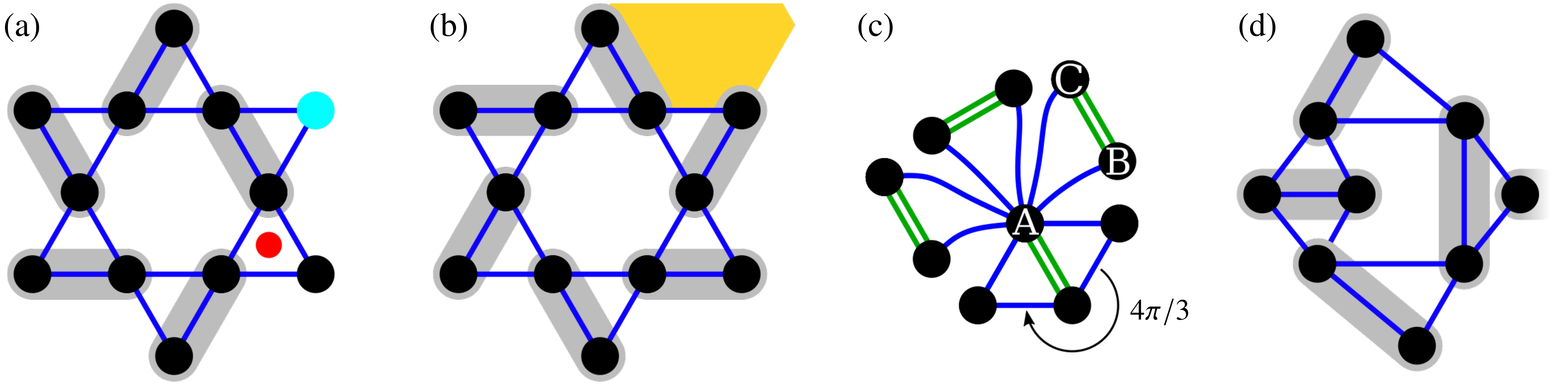}
    \caption{For a $n$-gon surrounded by triangles, (a) if we place a dimer on the edge of the $n$-gon, we seclude one vertex (colored in cyan) from the rest of the assembly. There can not be a situation where all the triangles contain a dimer (see the triangle with the red dot) and (b) even if we place the dimers on the surrounding triangles, we are placing dimers on the edges of the surrounding $n$-gons (see the yellow truncated hexagon), which would eventually lead to triangles with no dimers on the extended lattice. (c) The lattice must have an odd coordination number for a dimer state to exist. (d) The first half of Theorem 1 does not apply to a lattice containing irregular polygons. However, the latter half of the theorem is still valid.}
    \label{fig:fig5}
\end{figure}

\textbf{Definition 2:} $n$-gon -- A regular polygon with $n$ sides, where $n>3$ is finite.

\textbf{Theorem 1:} An exact dimer state on a uniform lattice is only possible if no dimer exists on an edge shared by a triangle and a $n$-gon. Thus, a lattice with purely corner-sharing triangles can not bear an exact dimer state. 

\emph{Proof:} 
For an exact dimer state, one must write all the interactions in the Hamiltonian as interactions on a triangle. Therefore, every side of the $n$-gon must be part of a triangle. In other words, every $n$-gon in the lattice is surrounded by $n$ triangles. Now, on this assembly of a $n$-gon surrounded by $n$ triangles ($2n$ vertices in total), if we place a dimer on one side of the $n$-gon, it secludes one vertex, i.e., this vertex can no longer take part in forming a dimer with any of the other vertices in the assembly. So, we come to a situation where we are supposed to place $n-1$ dimer on the rest $n-1$-triangles, but we are left with only $2n-3$ vertices which can be a part of a dimer. A pictorial representation of this situation is given in Fig.~\ref{fig:fig5} (a). Fig.~\ref{fig:fig5} (b) shows a situation where the dimer configuration has been satisfied fully on the assembly of one $n$-gon and surrounding $n$ triangles, but there we have to place dimers on the edges of adjacent $n$-gons. Again, we will be led to a similar impossibility as above when we extend the lattice further. A lattice with purely corner-sharing triangle geometry is a tiling of triangle(s) and $n$-gon(s). Hence, any dimer there would always reside on an edge shared by a triangle and a $n$-gon. Therefore, there can not exist an exact dimer state. One can readily see that such an argument is also true for 3D systems.

Note: a uniform 2D lattice made of triangles and $n$-gons is only possible if $n=4$, $6$, or $12$. 

\textbf{Corollary 1:} 
For an exact dimer state on a uniform lattice, all dimers must be shared between two triangles. 

\emph{Proof:} If all dimers do not reside on the edges shared by two triangles, this, in turn, means that at least one of them is located on an edge shared by a triangle and a $n$-gon. Therefore, this can not result in an exact dimer state.

\textbf{Lemma 1:} For a pure dimer state to exist on a uniform lattice, a necessary condition is to have a coordination number of $5$. 

\emph{Proof:} As a dimer must be supported by two triangles, each end of the dimers already have at least $3$ edges. If we now focus on one end of this dimer and assume that there is no other edge that connects to this end, that means the $n$-gon which is adjacent to both of these triangles has an internal angle $>\pi$ ($=4\pi/3$), which is not possible (refer to Fig.~\ref{fig:fig5} (c)). So, $n>3$. Now, if one new edge connects to the one end (say A) of this dimer, it must also connect to one end (say B) of another dimer. We call the other end of the second dimer C. Now, to write the interactions on these two new edges as the interactions on a triangle, there must be an edge that connects A and C (see Fig.~\ref{fig:fig5} (c)). Thus, we always have an odd coordination number greater than $3$. For uniform lattices, the possible coordination numbers are only between $3$, and $6$. Therefore, the only choice we are left with is the coordination number of $5$. 

\textbf{Corollary 2:} The Shastry-Sutherland lattice and the Maple-leaf lattice exhaust the list of the lattices with $1$-uniform semi-regular tilings that can host an exact dimer phase. 

\begin{figure}
    \centering
    \includegraphics[width=0.5\textwidth]{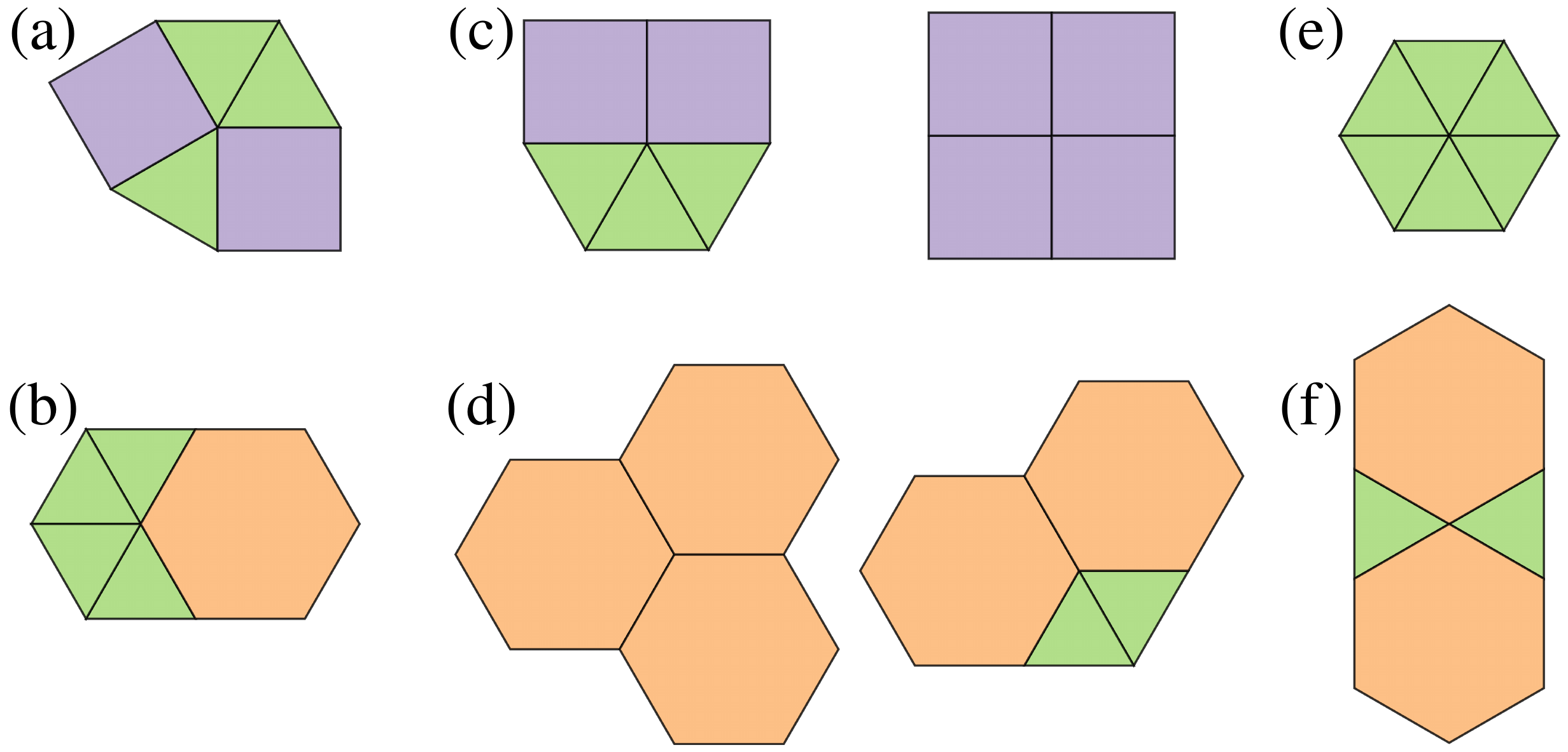}
    \caption{(a) and (b) are the unit tiles ($3.3.4.3.4$ and $3.3.3.3.6$ respectively) of the Shastry-Sutherland and the Maple-Leaf lattice, respectively. (c) and (d) are the other possible vertices formed by the tiles (a) and (b), respectively, where the edges are not part of a triangle. (e) and (f) are possible vertices with even coordination number. (e) can appear for both (a) and (b) tiling. (f) appears only for (b) tiling.}
    \label{fig:fig4}
\end{figure}

\emph{Proof:} Of the $21$ semi-regular tilings, only $2$ (refer to Fig.~\ref{fig:fig4} (a) and (b)) have a coordination number $5$ and where all the edges are part of a triangle. The $1$-uniform tilings of the former generate the Shastry-Sutherland lattice, and the latter gives rise to the Maple-leaf lattice. Therefore, the Shastry-Sutherland and the Maple-leaf lattices are the only two lattices of $1$-uniform semi-regular tilings that host an exact dimer state. 

We want to make a minor comment here that apart from these $21$ semi-regular tilings, there is another trivial tiling -- $3$-triangles and an $\infty$-gon. This tiling generates the Majumdar-Ghosh model.

\textbf{Lemma 2:} 
The Shastry-Sutherland lattice and the Maple-leaf lattice exhaust the list of all the 2D lattices with uniform semi-regular tilings that can host an exact dimer phase. 

\emph{Proof:} There are infinitely many lattices with $k$-uniform tilings with $k\ge2$~\cite{Grunbaum1977}. By definition, the $k$-uniform tilings contain $k$ different vertex types. One can readily figure out that the other vertex types (apart from the ones shown in Fig.~\ref{fig:fig4} (a) and (b)) one can construct out of these two tiles will either have at least one edge that is not part of a triangle (Fig.~\ref{fig:fig4} (c) and (d)) or the vertex has an even coordination number (see Fig.~\ref{fig:fig4} (e) and (f)). An exact dimer state can not exist in both these situations. Therefore, apart from the Shastry-Sutherland lattice and the Maple-leaf lattice, no other nearest-neighbor 2D system with semi-regular tiling can bear an exact dimer ground state. 

Note that one can still find other uniform semi-regular tilings on the hyperbolic plane (e.g., snub tetrapentagonal tiling, snub pentapentagonal tiling, snub triheptagonal tiling, snub trioctagonal tiling, etc.), which can also host a pure dimer phase. Our geometrical arguments starting from Lemma 1 do not work there. Also, one can include irregular triangles and polygons in making the lattice, where the first part of Theorem 1 is not applicable. However, constructing the exact dimer state in such a situation would always mean at least one bond shared between two triangles. Thus, the second part of the theorem is a much stronger statement, and it always holds. An example of such an assembly of irregular triangles and polygons can be seen in Fig.~\ref{fig:fig5} (d). In these lattices, one needs a peculiar combination of spin-spin couplings to ensure an exact dimer ground state. Similar ideas have been used in the construction of the exact dimer states in Ref.~\cite{Siddharthan1999,SchmidtExact}, where the lattices are a mixture of corner-sharing and edge-sharing triangles, i.e., lattices of non-uniform tilings. Therein, the general scheme used to construct an exact dimer state is to couple systems with exact dimer states, e.g., the sawtooth chain or the maple-leaf unit, via a different unit containing an exact dimer state, e.g., the Shastry-Sutherland unit. 

\section{Details of the DMRG calculations}
\begin{figure}[h]
    \centering
    \includegraphics[width=0.65\textwidth]{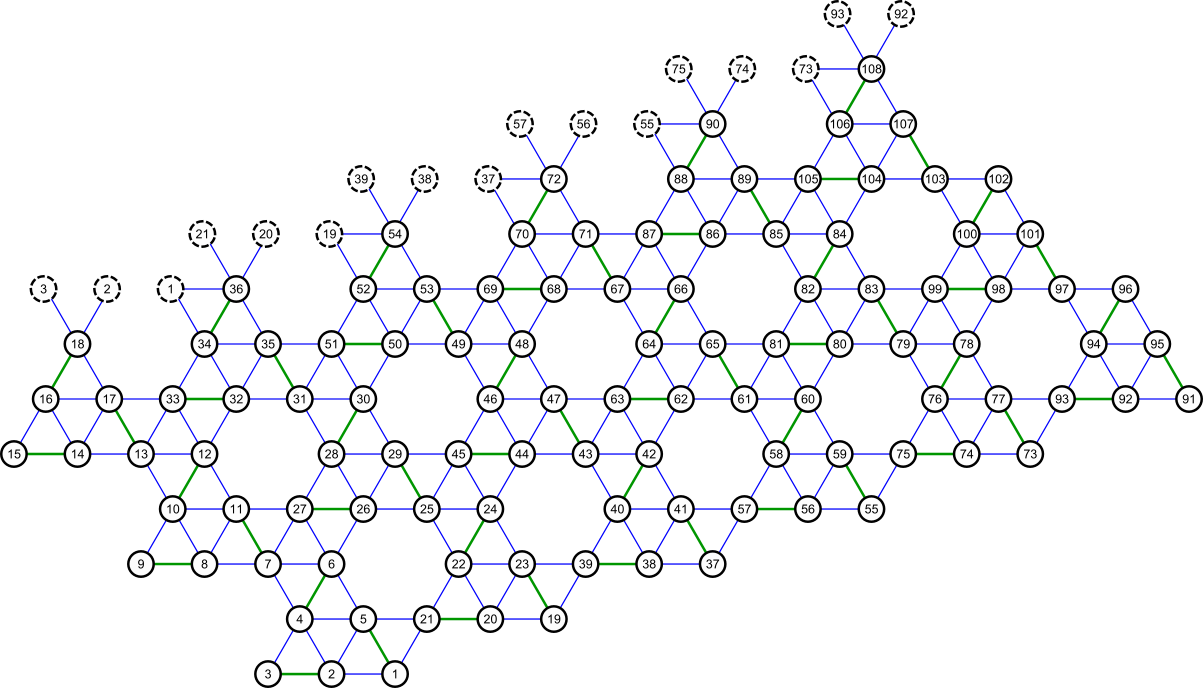}
    \caption{The $108$-site MLM cluster used to perform the DMRG calculations mentioned in the main text.}
    \label{fig:fig6}
\end{figure}
Our DMRG calculations are done for purely Heisenberg interactions with decreasing $\alpha$. The DMRG calculations are performed with the matrix product state (MPS) algorithm using the ITensor library \cite{Fishman2020} on a $108$ site cluster with mixed open and periodic boundary conditions. The $108$ site cluster is a $3\times6$ (along the $\vec{a}_1-\vec{a}_2$ and $\vec{a}_2$ directions, respectively) array of the unit-cell, with periodic boundary condition along the $\vec{a}_1-\vec{a}_2$ direction. The maximum bond dimension used for these calculations is $4096$. For each DMRG run we are performing 16 full sweeps. When the system assumes the exact dimer state, the energy of the system must be $-S(S+1)\alpha$ per spin. Our DMRG calculations find the same numerical value for the ground state energy for $\alpha>\alpha_c$, with a discarded entropy of zero, within the precision of our calculations ($10^{-16}$). A deviation from this energy value is interpreted as a departure from the exact dimer ground state. For $\alpha<\alpha_c$, the discarded entropy becomes $~10^{-6}$, which is also an indication of the disappearance of the pure nearest neighbor dimer correlations.

\section{Comments on further bond anisotropies on the model}
\begin{figure}[h]
    \centering
    \includegraphics[width=0.53\textwidth]{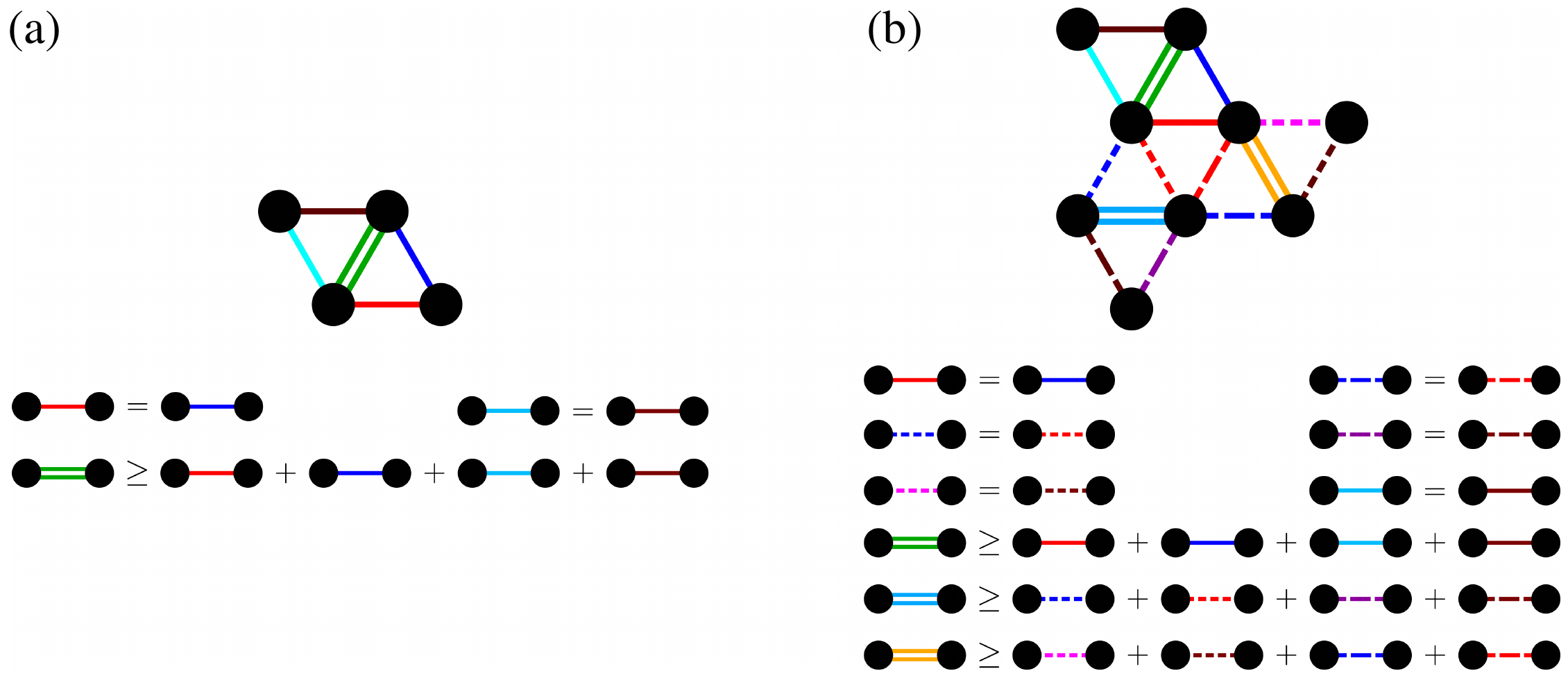}
    \caption{(a) The basic rhombus unit needed for an exact dimer state. The conditions for the exact dimer state are shown on the bottom.  (b) The full lattice made with three rhombus units with different couplings. This completely breaks down the $p6$ symmetry of the Maple-Leaf lattice. The conditions for the exact dimer state is also given.}
    \label{fig:fig6}
\end{figure}
The exact dimer state is robust against a variety of bond anisotropies, as long as a rather general condition is met. To realize this, we start from a rhombus with the dimer located on the short diagonal as a basic building block of the ex-D state, as shown in Fig.~\ref{fig:fig6}(a). 
This dimer configuration will lead to an ex-D state, as long as (i) both the bonds that form a triangle together with the diagonal bond have equal coupling strength and (ii) the diagonal coupling is at least as strong as the total coupling on all the other bonds.
From such rhombi, we can build up the whole lattice as shown in Fig~\ref{fig:fig6} (b), where the conditions are satisfied on all rhombi individually.
Note that this construction, in general, completely breaks the $p6$ symmetry of the Maple-Leaf lattice. A compound with a related structure has been reported in Ref.~\cite{Haraguchi2021} and modeled in Ref.~\cite{Makuta2021} where the $p6$ is only broken down to $p3$.
\bibliography{Refs}